# Detecting Bot Activity in the Ethereum Blockchain Network

Morit Zwang[1], Shahar Somin[1,2],
Alex 'Sandy' Pentland[2], and Yaniv Altshuler[1,2]

[1]Endor Ltd.
[2]MIT Media Lab, Cambridge, MA, USA
morit@endor.com
{shaharso, pentland, yanival}@media.mit.edu

## Introduction

The Ethereum blockchain network is a decentralized platform enabling smart contract execution and transactions of Ether (ETH) [1], its designated cryptocurrency. Ethereum is the second most popular cryptocurrency with a market cap of more than 100 billion USD, with hundreds of thousands of transactions executed daily by hundreds of thousands of unique wallets. Tens of thousands of those wallets are newly generated each day.

The Ethereum platform enables anyone to freely open multiple new wallets [2] free of charge (resulting in a large number of wallets that are controlled by the same entities). This attribute makes the Ethereum network a breeding space for activity by software robots (bots). The existence of bots is widespread in different digital technologies and there are various approaches to detect their activity such as rule-base, clustering, machine learning and more [3,4]. In this work we demonstrate how bot detection can be implemented using a network theory approach.

## A Network Theory Approach

Being a platform used for human interactions, the Ethereum network can be described and modeled by a Network Theory approach. The degree distribution of such networks, for example, often displays a power law distribution [5]. This phenomenon can also be observed when constructing a network that represents Ethereum transactions between wallets—where each wallet is a vertex and a transaction between two wallets is an edge. (Fig. 1)

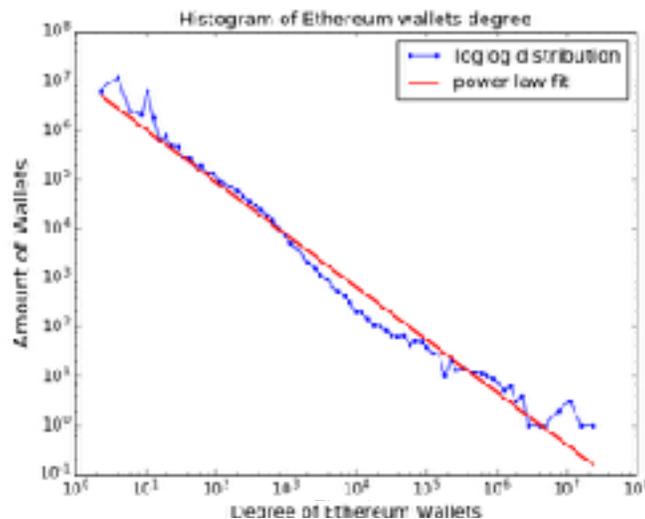

Previous research has demonstrated that time differences between consecutive events in many human activities display a power law distribution. This phenomenon is displayed in waiting time for call centers and e-mail communication [6] as well as in pausing time between transactions in a USD/DEM exchange [7]. In this work, we ask whether the time difference between consecutive Ethereum transactions demonstrates a power law distribution as well.

The time difference between consecutive transactions in this work refer to the number of minutes between every transaction and its prior transaction. The time difference was calculated for the transactions of each wallet separately, and we created a histogram from the time difference of all the transactions of all wallets in the Ethereum network. The histogram shows that, indeed, the time difference between consecutive transactions demonstrates a power law distribution. (Fig. 2)

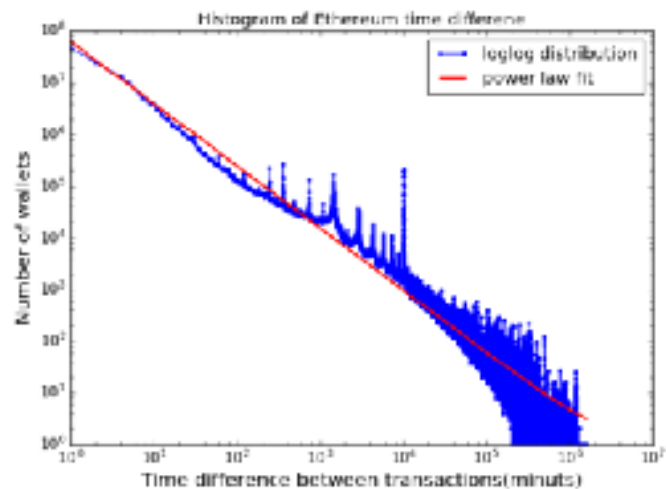

Every bin in the distribution contains a group of wallets which have made two consecutive transactions with the same time difference. It can be observed that the distribution of time differences between the consecutive transactions of all Ethereum wallets does not perfectly fit the power law model and is characterized by multiple spikes. Anomalies from the power law model in human behavior networks might represent the occurrence of potentially interesting events [8]. We can clearly distinguish between two types of anomalies:

**Periodic anomalies:** Anomalies consisting of a specific time difference which repeats itself in any random sample of a fixed time range. For example, when sampling a period of two days beginning at any random date and time, there will be a spike at the time difference of 24 hours. (Fig. 3a) The same spike will appear when sampling any random period of one week. In a one-week period, there are additional repeated spikes at a time difference of 48, 72, 96, 120 and 144 hours. (Fig. 3b) Analyzing the transactions which created these spikes reveals that many of the transactions were executed by mining pools distributing mining reward to pool members (mining pools are groups of users sharing their processing power in order to compete for the right to generate a block and win the mining reward [2]).

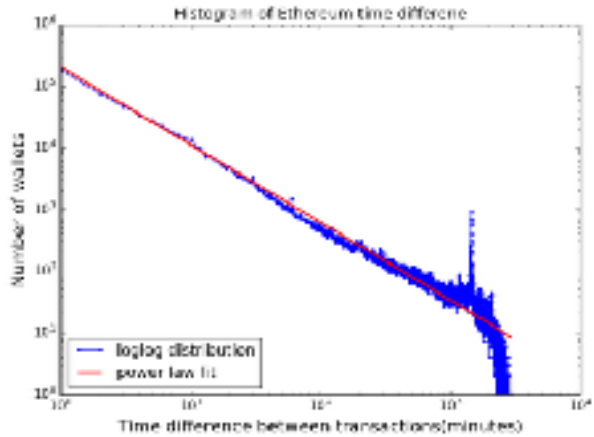 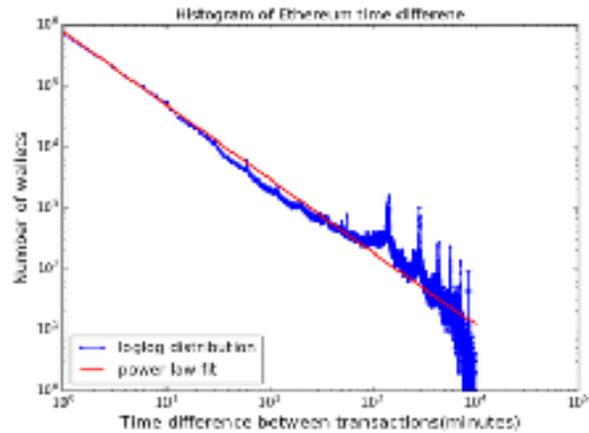

*Fig. 3a: An arbitrary two days sample*        *Fig. 3b: An arbitrary one-week sample*

**Irregular anomalies:** Anomalies consisting of a specific time difference, taking place only at a particular time period. For instance, displayed in Fig. 4 we can find a spike at a time difference of 1032 minutes (17 hours), observed on May 18, 2018 only. Analyzing the wallets whose transactions create these irregular spikes, we found that they took part in token Airdrops—a distribution of free tokens [9]. By creating many unique wallets and participating in an Airdrop, one entity can collect a large number of tokens. Such wallet activity is usually executed by bots. The presence of non-human activity provides an explanation for observed behavioral patterns which deviate from the power law model.

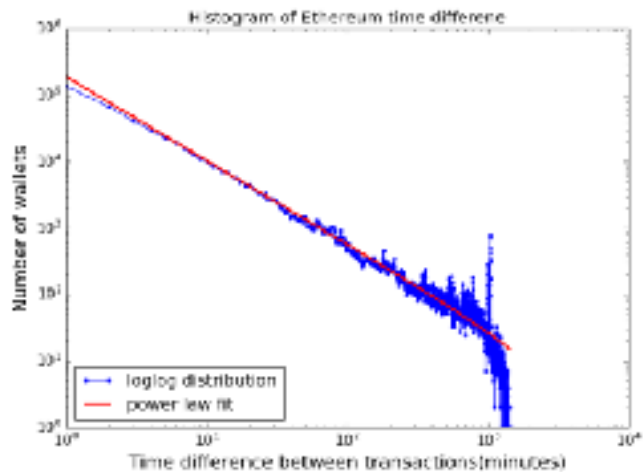

*Fig. 4 – One day sample at May 18, 2018. The spike is similar to the periodic spikes, but it appears at the time difference of 1032m only this specific day.*

## Conclusion

Each spike in both types of anomalies represents a collection of highly correlated wallets which deviate from the expected power law distribution rather than resembling spontaneous human activity. In some cases, anomalies from the power law model in human interaction networks may be evidence for emergency events [10]. In this case, we assume that transactions which are anomalous to the power law model represent non-human behavior executed by bots. This assumption is based on the nature of the anomalies (spikes occurring at a very specific time difference) and on the observation of other properties common to the anomalous transactions, such as having the same transaction value or the same destination wallet. Using a network theory approach, analyzing the distribution of time differences between consecutive transactions enables us to detect non-human activity.